\documentstyle[12pt]{article}
\begin{document}

\begin{titlepage}
\vspace{.3cm}
\begin{center} {\large \bf
SOFTLY BROKEN FINITE SUPERSYMMETRIC \\ GRAND UNIFIED THEORY}

\vglue 1.3cm

{\bf D.I.Kazakov, M.Yu.Kalmykov, I.N.Kondrashuk}

\vglue .3cm

{\it Bogoliubov Laboratory of Theoretical Physics,
Joint Institute for Nuclear Research, 141 980 Dubna, Moscow
Region, RUSSIA \\}

\vglue 0.3cm {\bf and} \vglue 0.3cm

{\bf A.V.Gladyshev}
\vglue .3cm

{\it Physics Department, Moscow State
University, 119 899 Moscow, RUSSIA}
\end{center}

\vspace{.3cm}

\begin{abstract}
 In the context of the standard SUSY GUT scenario, we present a
detailed analysis of the softly broken finite supersymmetric grand
unified theory. The model, albeit non-minimal, remains very rigid due
to the requirement of finiteness. It is  based on the $SU(5)$ gauge
group and is UV finite to all orders of perturbation theory. It
contains three generations of the matter fields together with four
pairs of Higgses. The requirement of UV finiteness fixes all the Yukawa
couplings at the GUT scale. Imposing the condition of universality on
the soft couplings at the Planck scale and then extending the condition
of finiteness to them, one gets a completely finite unified theory
above $M_{GUT}.$ This makes the fine-tuning procedure more meaningful
and leads to the usual Minimal Supersymmetric Standard Model below
$M_{GUT}$. All the  masses of the ordinary particles including Higgses
are obtained due to the Higgs mechanism at the electro-weak scale.  The
hierarchy of quark and lepton masses is related to that of v.e.v.'s of
 the Higgs fields and is governed by the Higgs mixing matrix in the
generation space.  Superpartners develop their masses according to the
RG equations starting from the soft terms at the Planck scale.
The suggestion of complete finiteness and maximal simplicity
of the unified theory leads to  the connection between the initial
values of soft SUSY breaking parameters, namely $m_0^2=1/3m_{1/2}^2,\
A_t=A_b=A_\tau=-m_{1/2}, \ B=-m_{1/2}$, so that the number of free
parameters is less than that of the MSSM.
\end{abstract}
\vspace{.5cm}
\hrule
{\footnotesize
\begin{tabbing}
e-mail: \= kazakovd@thsun1.jinr.dubna.su \\
\>  kalmykov@thsun1.jinr.dubna.su \\
\>  ikond@thsun1.jinr.dubna.su \\
\>  gladysh@thsun1.jinr.dubna.su
\end{tabbing} }

\end{titlepage}

\section{Introduction}

During the past few there has been a considerable interest in the
Minimal Supersymmetric Standard Model~\cite{MSSM} and in SUSY
GUTs~\cite{GUT}.  It is because of the remarkable unification of the
gauge couplings in these models~ \cite{ABF}, which leads to predictions
of the SUSY spectrum in the energy region within the reach of future
accelerators~\cite{Spectr}.  The detailed analysis performed by various
groups~\cite{Many,WE} is based on the SUSY GUT scenario with soft
supersymmetry breaking due to the supergravity mechanism and is
different only in details. It takes into account two-loop
renormalization group equations, one-loop corrections to the Higgs
potential, the heavy and light threshold effects and various
experimental constraints. Perhaps the most remarkable fact is that all
the requirements can be fulfilled simultaneously and are consistent with
very few free parameters. The predicted mass spectrum is concentrated in the
$10^2-10^3$ GeV region and is not very much model dependent. This leads to
the conclusion that the MSSM and SUSY GUTs provide us with a very promising
scenario that can be checked experimentally.

Of course, several problems remain unsolved. Besides the unknown explicit
mechanism of SUSY breaking parametrized by soft terms with five free
parameters~\cite{soft}, some problems of the Standard Model still
remain. For instance, the quark mass spectrum and the mixing of the
generations remain the biggest puzzles. Though some progress has
been made in these directions, there is no well accepted
solution. One of the most interesting attempts of this kind is the
one discussed in Ref.~\cite{Textures} where the values of the Yukawa
couplings and the Kobayashi-Maskawa mixing matrix at the
unification scale are given in the form of the so-called {\em
textures} which evolve to the observed values at low energies. The
textures themselves are chosen for reasons of maximal simplicity and
symmetry while the required parameters are fitted. The related idea
explores the possibility of determining the
Yukawa couplings by the infrared stable fixed point of the
theory lying beyond the Standard Model~\cite{fixedp}.

Another approach is based on a wider symmetry like $SO(10)$~\cite{SO10}.
In this case the masses of the heaviest generation arise from a single
renormalizable Yukawa interaction while the lighter masses are
generated by nonrenormalizable operators of the Grand unified theory.

There are naturally many attempts to consider some
non-minimal models that provide wider possibilities. Among them is the
so-called Next-to-minimal SSM~\cite{NMSSM} that allows one to relate
some soft breaking terms to the vacuum expectation value of the singlet
Higgs field.  However, this model does not touch the problems of the
quark mass spectrum and flavour mixing mentioned above.

Without denying these possibilities, we would like to
suggest an alternative approach that naturally arises in attempts to
construct SUSY GUTs, free from ultraviolet divergences~\cite{4,5,6}.

In the standard minimal SUSY GUT scenario, the theory possesses both the
supersymmetry and the unified gauge symmetry at the unification scale
with soft SUSY breaking terms arising from supergravity. At this scale
all quarks and leptons are massless and all their superpartners have
the same mass. Going down to lower energies the superpartners' masses
run according to the RG equations, split due to different interactions
and, thus, give us the mass spectrum. This is accompanied also by the
radiative spontaneous symmetry breaking, which leads to the
reconstruction of the vacuum state.  The latter, according to the usual
Higgs mechanism, provides us with masses for quarks, leptons and
$SU(2)$ gauge bosons and additional mass terms to their superpartners.

Since the Standard Model exploits the minimal version of the Higgs
mechanism with only one Higgs doublet to provide masses to all quarks and
leptons simultaneously, the mass spectrum is given by that of the Yukawa
couplings. In the MSSM one needs at least two doublets.
One doublet provides masses to up quarks; while the other, to down
quarks and leptons. Thus, we have two vacuum expectation values and their
ratio $ \tan\beta \equiv v_2/v_1$ is a free parameter in the model.
It is usually fitted from the experimental constraints. On the
other hand, the value of $\tan\beta$  can be found from the minimization of
the potential for neutral Higgses, if the parameters are known, and differs
>from unity. Thus, we can get a hierarchy if the potential has an asymmetric
minimum~\cite{7}, though it is not essential in the case when the
Yukawa couplings remain arbitrary.

This is not the case, however, in finite SUSY GUT models where
the Yukawa couplings at the GUT scale can be calculated and appear to be
degenerate with respect to generations. On the contrary, the number of
Higgs doublets increases, each being adjusted to a certain
flavour so that the mass spectrum is given by the spectrum of the
v.e.v.s of the Higgs fields rather than by that of the Yukawa
couplings.

The finite models, though non-minimal, still remain
almost as rigid as the minimal one and are distinguished by their ultraviolet
properties being absolutely UV finite to all orders of perturbation theory
\cite{4,5}. Let us remind the main properties of a finite SUSY GUT:

\begin{itemize}
\item  the number of generations is fixed by the requirement of finiteness,

\item  the representations and the number of the Higgs fields are fixed,

\item  all the Yukawa couplings are expressed in terms of the gauge one,

\item  various realistic possibilities are given by
$SU(5),SU(6),SO(10)$ and $E(6)$ gauge groups with few generations. An
abelian subgroup is not allowed.
\end{itemize}

The other attractive feature of a finite model is that if the gauge
symmetry is not broken, the parameters, including the soft terms, are
not running. This means that the couplings, masses, etc, at the GUT
scale have some absolute values. If they are governed by some symmetry,
it does not matter whether we impose this symmetry at the GUT or at the
Planck scale.

It should be mentioned that the attractiveness of  UV finite
models without gravity is often called into question since,
being renormalizable, SUSY GUTs are quite satisfactory in the practical
sense.  However, the motivation for SUSY itself is mainly due to the
cancellation of quadratic divergences which allows one to preserve the
hierarchy of the Higgs masses in SUSY GUTs.  The finite model is the
next step in this direction where not only quadratic, but logarithmic
divergences also cancel.

Below we consider a particular finite SUSY GUT model that is based on
the $SU(5)$ gauge group and is one of the simplest models of this type
deviating only slightly from the minimal SUSY GUT. It should be
stressed that this model is almost unique among possible finite
models, if one requires spontaneous symmetry breaking to take place
via the Higgs mechanism with elementary Higgs fields. The other
possible choice is $SU(6)$, but here too, the symmetry breaking takes
place along the $SU(5)$ pattern. Higher groups inevitably
explore composite Higgs fields \cite{6}.

On the other hand, if one accepts $SU(5)$, the number of generations is
exactly three without any other option. The singlets are not allowed
due to finiteness; hence the right handed neutrino is excluded. Thus,
the finiteness hypothesis happens to be very rigid and provides us with
a unique selection of a possible GUT distinguished by its mathematical
properties.

The paper is organized as follows. Sect.2 is devoted to a
general review of the $SU(5)$ supersymmetric finite unified theory. We
consider the simplest R-symmetric and B-L conserving superpotential and give
an explicit solution to the conditions of one-loop finiteness for the
Yukawa couplings. The soft SUSY breaking is considered in Sect.3. Going
along the same line we suppose that the soft SUSY breaking terms are
also finite above the GUT threshold which leads to the universality
condition at the Planck scale with some of the soft parameters being
fixed.  In Sect.4, the spontaneous breaking of SU(5) is discussed.  The
fine-tuning procedure which reduces the unified model to the MSSM
below $M_{GUT}$, is proposed. At the first step we are left with three
pairs of Higgs doublets, one to each generation. They develop
vacuum expectation values defining the Yukawa couplings of the low
energy theory. Then, at the next step, minimizing the Higgs potential
we separate the light pair of Higgses identified with that of the MSSM.
Heavy fields decouple at high energies.  In Sect.5, we analyze the
compatibility of our model with various experimental constraints such
as the values of the heavy quark masses, the proton lifetime, absence
of the flavour changing neutral currents, etc.  Finally, in Sect.6, the
main attractive features of our model and its general status are
summarized. The Appendix  contains the derivation of the   solution to
the Higgs potential minimization conditions.

\section{Unified Finite Theory}

The model is a supersymmetric $SU(5)$ gauge theory whose field
content and interactions are completely  defined by the requirement
of UV finiteness. From this point of view the finite model is even more
rigid than the minimal one. The SUSY breaking is achieved via the
supergravity mechanism in a usual way; but, the enlarged Higgs
sector requires more parameters.

\clearpage

\noindent \underline{Field Content}
$$
\begin{array}{llcl}
\mbox{Matter fields}: & \Psi _i-\overline{5} & \qquad \mbox{of} \ SU(5) & \
\ i=1,2,3 \qquad -
\mbox{generations} \\  & \Lambda _i - 10 & \qquad - / - &  \\
\mbox{Higgs fields}: & \Phi _{a} - 5 & \qquad - / - & \ \ a=1,2,3,4 \\
& \overline{\Phi}_a - \overline{5} & \qquad - / - &  \\
& \Sigma - 24 & \qquad - / - &
\end{array}
$$
\underline{Lagrangian}
\begin{eqnarray}
{\cal L}&=&{\cal L}_{SUSY}+{\cal L}_{Breaking},\nonumber  \\
{\cal L}_{SUSY}&=&{\cal L}_{Gauge}+{\cal L}_{Yukawa}+{\cal L}_{Mass},
\end{eqnarray}
where
\begin{eqnarray}
\displaystyle {\cal L}_{Yukawa}&=&y_1\Psi _iK_{ij}\overline{\Phi }_i\Lambda
_j+y_1^{\prime}\Psi _i\overline{\Phi }_4\Lambda _i+\frac{y_2}8\Phi _i\Lambda
_i\Lambda _i+\frac{y_2^{\prime}}8\Phi _4\Lambda _i\Lambda _i \nonumber \\
&& +y_3\overline{\Phi }_iS_{ij}\Sigma \Phi _j+y_3^{\prime}
\overline{\Phi }_4\Sigma \Phi _4+\frac{y_4}3\Sigma ^3, \label{YU}
\end{eqnarray} and
\begin{equation}
\label{Mass}\displaystyle {\cal L}_{Mass}=\overline{\Phi }_iM_{ij}\Phi _j+
\overline{\Phi }_4M\Phi _4+ \frac{M_0}2\Sigma ^2.
\end{equation}
Here the matrices $K$ and $S$ are unitary:
$$
K^{+}K=1,\ \ S^{+}S=1,
$$
$K$ being the Cabbibo-Kobayashi-Maskawa mixing matrix and $S$ playing the
same role in the Higgs sector. As we show below (Sect.4), the matrix
$S$ contains all information about the quark masses hierarchy at the
GUT scale.

\vglue 0.5cm

\underline{Yukawa couplings}: The requirement of UV finiteness is
already attained at the one-loop level. Besides the field content of
the model it defines also the superfield Yukawa couplings in terms of
the gauge one:
$$
\displaystyle Y_i\equiv \frac{y_i^2}{16\pi ^2}=c_i\frac{g^2}{16\pi ^2}
\equiv c_i\widetilde{\alpha}_G .
$$
For the Lagrangian (\ref{YU}), the RG equations for the Yukawa
couplings are~\cite{4,5}:
\begin{eqnarray}
\frac{dY_1}{dt}&=&Y_1\Bigl[
10Y_1+ 6Y_1^{\prime}+3Y_2+3Y_2^{\prime}+\frac{24}{5}Y_3
-\frac{42}{5}\tilde{\alpha}_G\Bigr] , \nonumber \\
\frac{dY_1'}{dt}&=&Y_1'\Bigl[
6Y_1+ 18Y_1^{\prime}+3Y_2+3Y_2^{\prime}+\frac{24}{5}Y_3
-\frac{42}{5}\tilde{\alpha}_G\Bigr] ,\nonumber \\
\frac{dY_2}{dt}&=&Y_2\Bigl[
4Y_1+ 4Y_1^{\prime}+9Y_2+6Y_2^{\prime}+\frac{24}{5}Y_3
-\frac{48}{5}\tilde{\alpha}_G\Bigr]  , \nonumber \\
\frac{dY_2'}{dt}&=&Y_2'\Bigl[
4Y_1+ 4Y_1^{\prime}+6Y_2+15Y_2^{\prime}+\frac{24}{5}Y_3
-\frac{48}{5}\tilde{\alpha}_G\Bigr]   ,  \\
\frac{dY_3}{dt}&=&Y_3\Bigl[
4Y_1+ 3Y_2 +\frac{63}{5}Y_3 +Y_3^{\prime}+\frac{21}{5}Y_4
-\frac{49}{5}\tilde{\alpha}_G\Bigr]  , \nonumber    \\
\frac{dY_3'}{dt}&=&Y_3'\Bigl[
12Y_1'+ 9Y_2' +3Y_3 +\frac{53}{5}Y_3^{\prime}+\frac{21}{5}Y_4
-\frac{49}{5}\tilde{\alpha}_G\Bigr]   , \nonumber      \\
\frac{dY_4}{dt}&=&Y_4\Bigl[
9Y_3 +3Y_3^{\prime}+\frac{63}{5}Y_4-15\tilde{\alpha}_G\Bigr].\nonumber
\end{eqnarray}
Here $t\equiv log(Q^2/M^2)$.

The finiteness solution contains one free parameter $c$:
$$
\begin{array}{llll}
c_1=c, & c_1^{\prime}=\frac 35-c, & c_2=\frac 43c, & c_2^{\prime}=\frac
43(\frac 35-c), \\
c_3=\frac{5}{6}(\frac 35 -c), & c_3^{\prime}=-\frac 52(\frac 25 -c), & c_4=
\frac{15}{14} & .
\end{array}
$$
\noindent
Since $c_i\geq 0,$ the parameter $c$ is restricted by
the inequality $\frac 25\leq c\leq \frac 35.$ In particular cases we
have:
$$ \begin{array}{llllllll} c=\frac 25, & c_1=\frac 25, &
c_1^{\prime}= \frac 15, & c_2=\frac 8{15}, & c_2^{\prime}=\frac 4{15},
& c_3=\frac 16, & c_3^{\prime}=0, & c_4= \frac{15}{14} ; \\ c=\frac 35,
& c_1=\frac 35, & c_1^{\prime }=0, & c_2=\frac 45, & c_2^{\prime }=0, &
c_3=0, & c_3^{\prime }=\frac 12, & c_4= \frac{15}{14}.  \end{array} $$

\noindent
In what follows, we take the case $c=\frac 25$. Recall that
$g_{SUSY}=\sqrt{2}g_{Non-SUSY}$. Later on we will use $g_{Non-SUSY} \equiv
g_G $ everywhere.

These relations for the Yukawa couplings are valid at one- and two-loop
levels and have to be corrected at higher loops~\cite{4}. The
corrections are finite and can be expressed either in terms of the
series in the renormalized gauge coupling, or in the regularization
parameter (for instance, $\varepsilon $ in dimensional regularization)
for the bare coupling~\cite{6}.

\section{Soft SUSY Breaking via Supergravity}

We accept a common procedure for the soft supersymmetry breaking via a
supergravity mechanism when supersymmetry is broken in a hidden sector
that couples to the observable world only via gravity.
All possible allowed soft supersymmetry breaking terms in a generic
$N=1$ SUSY theory have been analysed in Ref.~\cite{soft}.
It is usually assumed that they arise at the Planck scale and have
a universal form motivated by some supergravity models~\cite{sugra}.
To determine their evolution down to the GUT scale, one has to apply
the RG equations of a particular GUT model. In general, this may lead
to considerable splitting between mass parameters~\cite{Pol}, which
results in uncertainties in the low energy predictions.

In our case, since we want to construct a completely {\em finite}
GUT model, it is natural to determine the soft terms from
the requirement of finiteness. An arbitrary matrix structure of the
soft couplings has been considered in Ref.~\cite{Jones}, where it has
been shown that the requirement of one-loop finiteness defines the
trilinear and bilinear soft couplings to be proportional to the
corresponding Yukawa terms in a superpotential.
Therefore, we consider the following universal form
of ${\cal L}_{soft}$ at the Planck scale:
\begin{eqnarray} -{\cal L}_{soft}&=&
m_{\overline{\phi}}^2|\overline{\phi_i}|^2+m_{\phi}^2|\phi_i|^2+
m_{\overline{4}}^2|\overline{\phi_4}|^2+m_4^2|\phi_4|^2+
m_{\Sigma }^2|\Sigma|^2+m_5^2|\psi_i|^2+m_{10}^2|\lambda_i|^2
\nonumber \\
&&+\Bigl[B_{\Sigma }\frac{M_0}{2}\Sigma ^2+
B_{\phi}\overline{\phi }_iM_{ij}\phi _j+
B_4\overline{\phi }_4M\phi _4 \\
&&+A_1y_1\psi _iK_{ij}\overline{\phi }_i\lambda _j
+A_1^{\prime}y_1^{\prime}\psi _i\overline{\phi }_4\lambda _i
+A_2\frac{y_2}8\phi _i\lambda _i\lambda _i
+A_2^{\prime}\frac{y_2^{\prime}}8\phi _4\lambda _i\lambda _i \nonumber
\\ &&+A_3y_3\overline{\phi }_iS_{ij}\Sigma \phi _j
+A_3^{\prime}y_3^{\prime} \overline{\phi }_4\Sigma \phi _4
+A_4\frac{y_4}3\Sigma ^3 +
\frac{1}{2}M_5\lambda_{\alpha }\lambda_{\alpha } + h.c.
\Bigr] ,  \nonumber
\end{eqnarray}
where $\phi, \psi, \lambda,$ and $\Sigma$ are the scalar components of
the corresponding matter superfields and $\lambda_\alpha $ are the
gauginos.

The one-loop RG equations for the soft terms are:
\begin{eqnarray*}
\frac{dm^2_{10}}{dt}&=&\Bigl[
3Y_2(m^2_{\phi}+2m^2_{10}+A^2_2)
+3Y^{\prime}_2(m^2_{4}+2m^2_{10}+A_2^{\prime 2})   ,\\
&&+2Y_1(m^2_{\overline{\phi}}+m^2_{10}+m^2_{5}+A^2_1)
+2Y_1^{\prime}
(m^2_{\overline{4}}+m^2_{10}+m^2_{5}+A_1^{\prime 2})
-\frac{72}{5}\tilde{\alpha}_GM_{5}^2\Bigr] ,\\
\frac{dm^2_{5}}{dt}&=&\Bigl[
4Y_1(m^2_{\overline{\phi}}+m^2_{10}+m^2_{5}+A^2_1)
+4Y_1^{\prime}(m^2_{\overline{4}}+m^2_{10}+m^2_{5}+A_1^{\prime 2})
-\frac{48}{5}\tilde{\alpha}_GM_{5}^2\Bigr] ,\\
\frac{dm^2_{\Sigma}}{dt}&=&\Bigl[
\frac{21}{5}Y_4(3m^2_{\Sigma}+A^2_4)
+ 3Y_3(m^2_{\overline{\phi}}+m^2_{\phi}+m^2_{\Sigma}+A^2_3) ,\\
&&+Y_3^{\prime}
(m^2_{\overline{4}}+m^2_{4}+m^2_{\Sigma}+A_3^{\prime 2})
-20\tilde{\alpha}_GM_{5}^2\Bigr],\\
\frac{dm^2_{\overline{\phi}}}{dt}&=& \Bigl[
4Y_1(m_{\overline{\phi}}^2+m_{10}^2+m_5^2+A_1^2)
+\frac{24}{5}Y_3(m_{\overline{\phi}}^2+m_{\phi}^2+m_{\Sigma}^2
+A_3^2)-\frac{48}{5}\tilde{\alpha}_GM_5^2\Bigr], \\
\frac{dm^2_{\phi}}{dt}&=& \Bigl[
3Y_2(m_{\phi}^2+2m_{10}^2+A_2^2)
+\frac{24}{5}Y_3(m_{\overline{\phi}}^2+m_{\phi}^2+m_{\Sigma}^2
+A_3^2)-\frac{48}{5}\tilde{\alpha}_GM_5^2\Bigr], \\
\frac{dm^2_{\overline{4}}}{dt}&=& \Bigl[
12Y_1'(m_{\overline{4}}^2+m_{10}^2+m_5^2+A_1^{\prime 2})
+\frac{24}{5}Y_3'(m_{\overline{4}}^2+m_{4}^2+m_{\Sigma}^2
+A_3^{\prime 2})-\frac{48}{5}\tilde{\alpha}_GM_5^2\Bigr] ,\\
\frac{dm^2_{4}}{dt}&=& \Bigl[
9Y_2'(m_{4}^2+2m_{10}^2+A_2^{\prime 2})
+\frac{24}{5}Y_3'(m_{\overline{4}}^2+m_{4}^2+m_{\Sigma}^2
+A_3^{\prime 2})-\frac{48}{5}\tilde{\alpha}_GM_5^2\Bigr] ,\\
\frac{dM}{dt}&=&M\Bigl[6Y_1'+\frac{9}{2}Y_2'+\frac{24}{5}Y_3'-
\frac{24}{5}\tilde{\alpha}_G\Bigr] ,\\
\frac{dM_{ij}}{dt}&=&M_{ij}\Bigl[2Y_1+\frac{3}{2}Y_2+\frac{24}{5}Y_3-
\frac{24}{5}\tilde{\alpha}_G\Bigr] ,\\
\frac{dM_0}{dt}&=&M_0\Bigl[3Y_3+Y_3'+\frac{21}{5}Y_4-10
\tilde{\alpha}_G\Bigr], \\
\frac{dM_5}{dt}&=&0 .
\end{eqnarray*}
The RGEs for the trilinear SSB parameters $A_i$ and quadratic terms
$B_i$ can be obtained from the RGEs of the corresponding Yukawa couplings
$Y_i$ and mass parameters by the replacement~\cite{Pol}
$$
\frac{dY_i}{dt}=Y_i\left[a_{ij}Y_j-b_i\tilde{\alpha}_G\right]\Rightarrow
\frac{dA_i}{dt}=\left[a_{ij}Y_jA_j+b_i\tilde{\alpha}_GM_5\right],
$$
$$
\frac{dM_i}{dt}=M_i\left[a_{ij}^{\prime}Y_j-b_i^{\prime}\tilde{\alpha}%
_G\right] \Rightarrow
\frac{dB_i}{dt}=2\left[a_{ij}^{\prime}Y_jA_j+b_i^{%
\prime}\tilde{\alpha}_GM_5\right],
$$
so that if $A_i=-M_5$, neither $A_i$ nor $B_i$ are running in the
finite model.

The condition of finiteness for the soft terms has the following solution:
\begin{eqnarray*}
m^2_{\overline{\phi}}&=&m^2_{\overline{4}}=m^2_{\overline{\phi}} ,\\
m^2_{\phi}&=&m^2_{4}=\frac{2}{3}M_5^2-m^2_{\overline{\phi}} ,\\
m^2_{10}&=&\frac{1}{6}M_5^2+\frac{1}{2}m^2_{\overline{\phi}}, \\
m^2_{5}&=&\frac{5}{6}M_5^2-\frac{3}{2}m^2_{\overline{\phi}} ,\\
m^2_{\Sigma}&=&\frac{1}{3}M_5^2,
\end{eqnarray*}
which is independent of the parameter $c$. If one assumes $m_5=m_{10}$,
one gets:
\begin{equation}
m^2_{\overline{\phi}}=m^2_{\phi}=m^2_{10}=m^2_{5}=m^2_{\Sigma}
=\frac{1}{3} M_5^2 .
\end{equation}

These one-loop conditions coincide with the general ones obtained in
Ref.\cite{Jones}. As has been shown by explicit
calculations~\cite{Jones2}, the one-loop finiteness of the soft terms
guarantees this property  at the two-loop level. Moreover, the
supergraph spurion mehtod~\cite{soft} allows one to expand this result
to an arbitrary number of loops~\cite{Yamada}. Indeed in
Ref.~\cite{Yamada}, the rules have been formulated how to get the
renormalization of the soft terms from the renormalization of the
Yukawa superpotential by a simple substitution, provided the
regularization prescription preserves supersymmetry.  Since we assume
that the regularization is SUSY invariant,  the finiteness of the soft
terms is valid in all orders of PT, if the one-loop conditions are
satisfied.

Thus, the requirement of finiteness naturally leads to the universality
of the soft breaking terms at the Planck/GUT scale.

The mass parameters $M_0, M_{ij}, M$ and $M_5$ are also not running.

\section{Reduction to the MSSM}

\subsection{Spontaneous breaking of SU(5)}

We follow the standard approach when the GUT symmetry is broken
spontaneously in a usual way by the vacuum expectation value of the Higgs
superfield $\Sigma$. For this purpose, we minimize the superpotential
$$
\displaystyle W_\Sigma =\frac{y_4}3\Sigma ^3+\frac{M_0}2\Sigma ^2,
$$
with the result
$$
\langle \Sigma \rangle =\left(
\begin{array}{ccccc}
V &  &  &  &  \\
& V &  &  &  \\
&  & V &  &  \\
&  &  & -\frac 32V &  \\
&  &  &  & -\frac 32V
\end{array}
\right),
$$
where $\displaystyle V\sim \frac{M_0}{y_4}\sim 10^{16}$ Gev.

After breaking of $SU(5)$ the $\Sigma $ field aquires the mass of the
order
of the GUT scale ($\sim 10^{16}$ Gev) and decouples, while the quintets
$\Phi $ and $\overline{\Phi }$ split into doublets and triplets.
Their mass terms look like
$$
y_3\overline{\Phi }_iS_{ij}\langle \Sigma \rangle \Phi _j+\overline{\Phi }
_iM_{ij}\Phi _j= \overline{\Phi _i}\left(
\begin{array}{cc}
y_3S_{ij}V+M_{ij} &  \\
& -\frac 32y_3S_{ij}V+M_{ij}
\end{array}
\right)\Phi _j ,
$$
and
$$ y_3^{\prime}\overline{\Phi }_4\langle
\Sigma \rangle \Phi _4+\overline{\Phi } _4M\Phi _4=\overline{\Phi
}_4\left( \begin{array}{cc} y_3^{\prime}V+M &  \\ & -\frac
32y_3^{\prime}V+M \end{array} \right) \Phi _4 . $$

In the first case, depending on the details of the fine tuning
procedure, we have several possibilities, namely, one can have both
the triplets and doublets to be heavy or one of them to be light
according to the choice of the matrices $S_{ij}$ and $M_{ij}$. In the
latter case, since $y_3^{\prime}=0$, there is no fine tuning and one
has both triplet and doublet to be heavy. All the heavy fields of the
theory decouple below the GUT scale.

The requirement of finiteness leads to the unitarity of the matrix
$S$. One can represent an arbitrary unitary matrix in the
following form:
\begin{eqnarray} S = \bar X\left( \begin{array}{ccc}
e^{i\theta_1} & 0 & 0 \\ 0  & e^{i\theta_2} & 0  \\ 0  & 0 &
e^{i\theta_3} \end{array} \right)X^T = \bar X D X^T, \ \ \bar X^T\bar X
= I, \ \ X^TX = I , \nonumber
\end{eqnarray}
where $X$ and $\bar X$ are
some real orthogonal matrices and $D$ is a unitary diagonal matrix. As
can be shown, one common phase can be absorbed into the redefinition of
the fields. Therefore, in what follows, we put $\theta_3 = 0.$

While the unitarity of S is dictated by finiteness, the mass matrix $M_{ij}$
is absolutely arbitrary.  Our choice of  $M$ is motivated by the
following requirements:
\begin{enumerate}
\item[i)] the presence of light Higgs doublets and decoupling of
the Higgs triplets;
\item[ii)] the absence of Goldstone bosons that may appear if the
continuous global flavour symmetry in the Higgs sector is spontaneously
broken;
\item[iii)] the reduction to the Standard Model at low energies.
\end{enumerate}
To fulfil these requirements we choose the matrix $M_{ij}$ in the form:
\begin{equation} M = \bar X (RI+T^{\prime}D)X^T,
 \label{m} \end{equation}
and perform the following fine-tuning procedure:
\begin{equation} T = T^{\prime} - \frac 32y_3V, \ \ R\sim T
 \sim V, \ \ R+T =\mu \sim 10^3 \ Gev .   \label{f}
\end{equation}

Since in the finite model none of the parameters is running
above the GUT scale, it is worth noticing that the fine-tuning
procedure here is more meaningful than in the other GUTs.

To argue that this choice of $M$ satisfies all the afore-mentioned
requirements, we analyze the theory below $M_{GUT}$ where $SU(5)$
is spontaneously broken. After decoupling of the heavy triplets, the
effective $SU(3)\times SU(2)\times U(1)$ invariant superpotential is:
\begin{eqnarray}
{\cal L}_{Yukawa}&=&\sqrt{\frac 25}g\Psi
_iK_{ij}\overline{\Phi }_i\Lambda _j+\frac 18\sqrt{\frac 8{15}}g\Phi
_i\Lambda _i\Lambda _i \Rightarrow \nonumber \\
&\Rightarrow &
\left(\sqrt{\frac 25}gQ_j^bK_{ij}\overline{H}_i^aD_i +\sqrt{\frac
25}gL_i^b\overline{H}_i^aE_i+\sqrt{\frac 8{15}}
gQ_i^bH_i^aU_i\right)\epsilon_{ab}, \label{Y} \\
{\cal L}_{Mass}&=&\overline{\Phi}_iM^{\prime}_{ij}\Phi_j \Rightarrow
\overline{H}_iM^{\prime}_{ij}H_j=
\overline{H}_i(\bar X(RI + TD)X^T)_{ij}H_j \; ,
\end{eqnarray}
where $M'=M-\frac{3}{2}y_3VS$, \ $a,b=1,2$ are the $SU(2)$ indices
and $\epsilon_{12}=1$.

Three pairs of the Higgs doublets have the following quantum numbers:
$$
\overline{H}_i(1,2,-\frac 12)=\left(
\begin{array}{c}
\bar H_i^0 \\
\bar H_i^{-}
\end{array}
\right) ,\ \ \ H_i(1,2,\frac 12)=\left(
\begin{array}{c}
H_i^{+} \\
H_i^0
\end{array}
\right) .
$$
The soft SUSY breaking terms below $M_{GUT}$ take the following form:
\begin{equation}
-{\cal L}_{Breaking}=m_0^2\sum\limits_i|\varphi
_i|^2+\frac{1}{2}\left(m_{1/2}\sum\limits_k \lambda _k \lambda _k
+ h.c. \right) \label{soft}
\end{equation}
$$ +\left(
A_Dy_D\widetilde{q}_j^bK_{ij}\overline{H}_i^a\widetilde{d}_i+
A_Ly_L\widetilde{l}_i^b\overline{H}_i^a\widetilde{e}_i
+A_Uy_U\widetilde{q}_i^bH_i^a\widetilde{u}_i+ B\overline{H}_i^a
M'_{ij}H_j^b + h.c. \right)\epsilon_{ab} ,
$$
where we have introduced the notation: $M_5=m_{1/2}$, $
\varphi_i$ are all the squark and slepton fields and $\lambda_k$ are the
gauginos.

The last term deserves special comment. It contains the mixing of the
Higgses in the generation space similar to the quark mixing via
the Kobayashi-Maskawa matrix $K$. This matrix will play the key role in
constructing the quark mass spectrum.

The boundary conditions at the GUT scale are:
\begin{equation}
m_0^2=\frac{1}{3}m_{1/2}^2, \ \ A_U=A_D=A_L=-m_{1/2}, \ \ B\equiv
B_{\Phi}=-m_{1/2} .
\end{equation}
The last equality follows from the fine-tuning requirement for the soft
terms at the GUT scale.

Therefore, we end up with the following set of free parameters:

\begin{itemize}
\item  3 gauge couplings $\alpha _i$,

\item  Mixing matrices $K_{ij}$ and $S_{ij}$,

\item  Mass terms $m_{1/2},R,T$.
\end{itemize}

\subsection{The Higgs Potential}

The tree level scalar Higgs potential consists of the SUSY part of the
Lagrangian and the soft terms

\begin{equation}
V(\bar{H}_i,H_i)= V_{SUSY}+V_{Soft},
\end{equation}
where
\begin{eqnarray}
V_{SUSY}&=&\overline{H}_i^*\overline{m}_{ij}\overline{H}_j + H_i^*m_{ij}H_j
+\frac{g^2+g^{\prime 2}}8(|\overline{H}_i|^2-|H_i|^2)^2 \nonumber \\
&&+\frac{g^2}4\left[|\overline{H}_i^{*}\overline{H}_j|^2
-|\overline{H}_i^{*}\overline{H}_i|^2
+|H_i^{*}H_j|^2-|H_i^{*}H_i|^2+2|\overline{H}_i^{*}H_j|^2\right]
\end{eqnarray}   with
$m_{ij}=(M^{\prime +}M^{\prime})_{ij} = (X((R^2+T^2)I +
RT(D^*+D))X^T)_{ij}$,
$\overline m_{ij}=(M^{\prime}M^{\prime +})_{ij}$ $ =(\bar X ((R^2+T^2)I
+ RT(D^*+D))\bar X^T)_{ij}$ and the soft terms are given by
eq(\ref{soft}).

Combining these equations, one obtains the following scalar potential:
\begin{eqnarray*}
V(\overline{H}_i,H_i) &=&  (m_{\bar\phi}^2 + R^2 +T^2)|\overline{H}_i|^2
+ RT\overline{H}_i^*\left(\bar X(D^* + D)\bar
X^T\right)_{ij}\overline{H}_j  \\
&+& (m_{\phi}^2 + R^2 + T^2)|H_i|^2 + RT H_i^*\left(X(D^*
+ D)X^T\right)_{ij}H_j \\
&+& B \left(\overline{H}^a_i\left(\bar X(RI +
TD)X^T\right)_{ij}H^b_j\epsilon_{ab} + h.c.\right)
+\frac{g^2+g'{}^2}8(|\overline{H}_i|^2-|H_i|^2)^2 \\ &+&
\frac{g^2}4\left[|\overline{H}_i^{*}\overline{H}_j|^2
-|\overline{H}_i^{*}\overline{H}_i|^2 +|H_i^{*}H_j|^2-|H_i^{*}H_i|^2
+2|\overline{H}_i^{*}H_j|^2\right]   .
\end{eqnarray*}

Due to our fine-tuning procedure, eqs.(\ref{m}),(\ref{f}), this
potential still contains the heavy Higgs fields with the masses of the
order of the GUT scale. To separate these states, we perform the
rotation in the Higgs sector and introduce the new fields  $H = X
H^{\prime}$, $\overline H = \bar X \overline H^{\prime}.$ Doing this,
one can rewrite the potential as
\begin{eqnarray}
V(\overline{H_i}^{\prime}, H_i^{\prime}) &=&
(m_{\bar\phi}^2 + R^2
+T^2)|\overline{H_i}^{\prime}|^2 + RT\overline{H_i}^{\prime *}
(D^* +D)_{ij}\overline{H_j}^{\prime} \nonumber\\
&+& (m_{\phi}^2 + R^2 + T^2)|H_i^{\prime}|^2
+ RT H_i^{\prime *}(D^* +D)_{ij}H_j^{\prime} \label{10}\\
&+&
B \left(\overline{H_i}^{\prime a} (RI + TD)_{ij}H_j^{\prime
b}\epsilon_{ab} + h.c.\right)
+\frac{g^2+g'{}^2}8(|\overline{H_i}^{\prime}|^2-|H_i^{\prime}|^2)^2
\nonumber\\
&+& \frac{g^2}4\left[|\overline{H_i}^{\prime
*}\overline{H_j}^{\prime}|^2 -|\overline{H_i}^{\prime
*}\overline{H_i}^{\prime}|^2 +|H_i^{\prime
*}H_j^{\prime}|^2-|H_i^{\prime *}H_i^{\prime}|^2
+2|\overline{H_i}^{\prime *}H_j^{\prime}|^2\right]. \nonumber
\end{eqnarray}

The potential (\ref{10}) is a simple generalization of
the MSSM \cite{MSSM} but differs from the latter by the extension of
the Higgs sector.  The electroweak symmetry breaking and the Higgs
sector of the broken theory in the models with  the Higgs potential of
this type have been analyzed in detail in Ref.~\cite{IK}.
This  potential is positive definite and has no
minima different from zero at the GUT scale due to supersymmetry like
in the MSSM. However, it develops the non-trivial minima radiatively,
thus leading to the radiatively induced spontaneous breaking of the
electroweak symmetry, just like in the standard scenario.  The
parameters of the potential evolve to the low energy values
according to the renormalization group equations.

When evolving to low energies the relations between different
parameters of the potential change, and, under some conditions,
the Higgs fields  gain nonzero v.e.v's. From the physical point of view,
we are interested in the minima that are achieved on the Higgs
field configurations that are gauge equivalent to the neutral real
ones, namely $ < \overline{H}'_i>=U\left( \begin{array}{c}
\bar v_i \\ 0
\end{array} \right) ,<H_i'>$ $=U\left(
\begin{array}{c} 0 \\
v_i \end{array}
\right), $ where $U$ is some $SU(2)$ matrix.
 In this case, the tree level minimization equations take the
following form:
\begin{eqnarray}
\frac 12\frac{\delta V}{\delta
\overline{H_i}}&=&{{\cal M}_1^2}_{ij}\overline{v}_j+
B{\cal M'}_{ij}v_j+\frac{g^2+g^{\prime
}{}^2}4(\overline{v}_k^2-v_k^2)\overline{v}_i=0, \nonumber\\
& & \label{mini}             \\
\frac 12\frac{\delta V}{\delta
H_j}&=&v_i{{\cal M}_2^2}_{ij}+\overline{v}_iB{ \cal M'} _{ij}-\frac{
g^2+g^{\prime }{}^2}4(\overline{v}_k^2-v_k^2)v_j=0, \nonumber
\end{eqnarray}
where
{\small
$$ {\cal M}_1^2=\left(
\begin{array}{ccc}
m_{\bar\phi}^2+R^2+T^2+2RT\cos\theta_1 &0&0 \\
0&m_{\bar\phi}^2+R^2+T^2+2RT\cos\theta_2& 0 \\
0& 0&m_{\bar\phi}^2+(R+T)^2 \end{array}\right),
$$

$$
{\cal M}_2^2=\left(
\begin{array}{ccc}
m_{\phi}^2+R^2+T^2+2RT\cos\theta_1 &0&0
\\ 0&m_{\phi}^2+R^2+T^2+2RT\cos\theta_2& 0 \\
0& 0& m_{\phi}^2 + (R+T)^2 \end{array}\right),
$$

$$
{\cal M'}=\left(
\begin{array}{ccc}R+ T\cos\theta_1 & 0&0 \\
 0 & R+ T\cos\theta_2 &  0 \\
0& 0& R+T \end{array}\right).
$$ }

The Yukawa part of the superpotential expressed in terms of the new
Higgs fields looks like:
$${\cal L}_{Yukawa}=
\left(y_D Q_j^b K_{ij} {\bar X}_{ik} \overline{H}_k'^a D_i
+ y_LL_i^b {\bar X}_{ik}\overline{H}_k'^a E_i +
y_U Q_i^b X_{ik} H_k'^a U_i\right)\epsilon_{ab}.$$
Due to the fine-tuning convention (\ref{m}),(\ref{f}) the only real and
positive solution of eqs.(\ref{mini}) which gives the v.e.v's of an
order of $M_Z$ is (the details are given in the Appendix):
\begin{eqnarray*} v_i = \left(\begin{array}{c}  0 \\
0  \\ u \end{array}\right), \ \ \overline v_i = \left(\begin{array}{c}
0 \\ 0 \\ \overline u \end{array}\right). \end{eqnarray*}
Here $u$ and  $\overline u$ are the same as $v_1$ and $v_2$ of the MSSM
\begin{eqnarray*}
u &=& \sqrt{\left(m_1^2 + m_2^2 \pm \sqrt{(m_1^2 +
m_2^2)^2 - 4B^2\mu^2} \right)F_{\pm}(\mbox{ $B^2\mu^2$})},
\label{u} \\ \bar u &=& -\mbox{{
sign($\mu$})}\sqrt{\left(m_1^2 + m_2^2 \mp \sqrt{(m_1^2 + m_2^2)^2 -
4B^2\mu^2}\right)F_{\pm}(\mbox{ $B^2\mu^2 $})}, \label{baru}
\end{eqnarray*}
where
$$ F_{\pm }(B^2\mu^2)=\frac 1{g^2+g^{\prime
2}}\frac{\pm (m_1^2-m_2^2)-\sqrt{ (m_1^2+m_2^2)^2-4B^2\mu
^2}}{\sqrt{(m_1^2+m_2^2)^2 -4B^2\mu^2}} $$
and $m_1^2=m_{\overline{\phi }}^2+\mu ^2,\ \ m_2^2=m_\phi ^2+\mu ^2, \
\mu=R+T.$ One takes ''+'' sign when $m_1^2>m_2^2$ and ''--'' sign  in
the opposite case. $u$ and $\bar u$ obey the usual equations of the MSSM
\begin{equation}
\bar u=v\cos \beta ,\ \ u=v\sin \beta ,\ \ v^2=\frac 4{g^2+g^{\prime 2}}
\frac{m_1^2-m_2^2\tan {}^2\beta }{\tan {}^2\beta -1},\ \ \sin 2\beta =-2
\frac{B\mu}{m_1^2+m_2^2}.
\end{equation}

For this minimum the Higgs doublets $\bar H'_3$ and $H'_3$ remain
light and are associated with the usual fields $H_1$ and $H_2$ of the
MSSM~\cite{MSSM}.  As for the doublets $\bar H'_{1,2}$ and $H'_{1,2}$,
due to our fine-tuning procedure (\ref{m}),(\ref{f}) they obtain masses
of order of $M_{GUT}$ and decouple below the GUT scale. This is also
true for their superpartners.

Re-expressing the original superfields $H_i$ and $\bar H_i$ through
$H'_i$ and $\bar H'_i$:
$$ H_i= X_{i1}H'_1+X_{i2}H'_2+X_{i3}H'_3, \ \
\bar H_i=\bar X_{i1}\bar H'_1+\bar X_{i2}\bar H'_2+\bar X_{i3}\bar
H'_3$$ and discarding the heavy first two terms, we obtain the MSSM with
$$H_1=\bar H'_3, \ \ H_2= H'_3 $$
and the usual Yukawa potential
\begin{equation}
{\cal L}_{Yukawa}=
\left(y_D\bar n_iQ_j^bK_{ij}H_1^aD_i+y_L\bar n_iL_i^b H_1^a
E_i + y_Un_iQ_i^bH_2^aU_i\right)\epsilon_{ab},
\end{equation}
where
$$\bar n_i = \bar X_{i3}, \ \ n_i=X_{i3}, \ \ \bar n_i^2=1,\ \ n_i^2=1.
$$
Due to the degeneracy of the Yukawa couplings the
Higgs v.e.v's play the key role in the creation of the quark and lepton
mass spectrum. As it appears, the hierarchy of the up quark masses is
defined by the vector $n_i,$ while the down quark hierarchy is defined
by $\bar n_i$. After decoupling of the heavy Higgs fields we end up
with the Minimal Supersymmetric Standard Model, where the Yukawa
couplings are given by the vacuum expectation values of the Higgs
fields
\begin{equation}
 y^U_i=n_iy^U, \ \ y^D_i=\bar{n}_iy^D, \ \ y^L_i=\bar{n}_iy^L, \label{in}
\end{equation}
and $y^U=\frac{4}{\sqrt{15}}g_{GUT}, \ y^D=y^L=\frac{2}{\sqrt{5}}g_{GUT}$
at the GUT scale.

As usual, the  quark and lepton masses are defined as eigenvalues of the
corresponding mass matrices at low energies. To find them one has to
run down the Yukawa matrices taking into account the initial conditions
(\ref{in}) and the generation mixing due to the Kobayashi-Maskawa
matrix $K$. As a result, one gets new nondiagonal Yukawa matrices which
again  have  to be diagonalized to extract the eigenvalues and the low
energy Kobayashi-Maskawa mixing matrix.

  We would like to stress that in this procedure all the initial information
about the quark mass hierarchy is contained in the Higgs sector of the finite
unified theory, namely, in the unitary Higgs mixing matrix $S$.

\section{Experimental Constraints}

The finite model described above appears to be a mathematically very
rigid one.  To argue its viability, we perform a brief analysis of the
compatibility of this model with existing experimental constraints;
namely, the unification of the gauge couplings, heavy quark and lepton
masses, the lower experimental limit on the proton lifetime and the
absence of the flavour changing neutral currents (FCNC).

\vglue 0.5cm

\underline{Unification of the gauge couplings}

Due to the heaviness of all the extra particles in the Finite model compared
to the MSSM and the reduction of the Finite model to the MSSM at low
energies, the unification of the gauge couplings takes place exactly in
the same manner as in Ref.~\cite{ABF}.  The numerical analysis is close
to that of the MSSM with large $\tan \beta $ and the RG equations
have exactly the same form as in MSSM~\cite{Many,WE}.  The only
difference is that due to the finiteness requirement the initial values
of the Yukawa couplings are fixed and the soft terms are more
restricted compared to the MSSM and hence one has less freedom to
fulfill all the requirements simultaneously.

\vglue 0.5cm

\underline{Heavy quarks and lepton masses}

One of the motivations for the Finite model has been to put some
impact on the quark spectrum. Contrary to the MSSM where the
Yukawa couplings are absolutely arbitrary, in our case, they are fixed
at the GUT scale. So, using the RG equations for the Yukawa matrices,
which coincide with those of the MSSM, and taking into account the
necessary thresholds, one can get their values at lower energies.
This would allow one to predict the values of the quark and lepton
masses.

Unfortunately, the requirement of finiteness does not fix all
the arbitrariness in the Yukawa matrices. For instance, the
Kobayashi-Maskawa matrix $K$ and the vectors $n_i$ and $\bar n_i$
remain arbitrary.  Practically, one can make some predictions only for
the third generation since the vectors $n_i$ and $\bar n_i$ are aligned
almost along the third axis in the generation space and one can choose
$n_i=\bar n_i=(0,0,1)$ in the first approximation,
and ignore the light generations.  Adjusting the soft parameters and
taking into account the light thresholds, one can arrive at the
experimental values of heavy quark and lepton masses. In doing this,
one has to take into account the difference between the running and the
pole masses~\cite{qm}.  Since the starting values of the bottom and top
Yukawa couplings, namely $y^D$ and $y^U$, are close to
each other, the large value of $\tan \beta $ is needed to explain the
big difference between the top and bottom masses.  The difference
between the bottom quark and $\tau $-lepton masses is due to the
different renormalization factors.

Prediction of the top quark mass in the finite model has first been
considered in detail in Ref.~\cite{Zoupanos}.  Following the arguments
analogous to those of the present paper, Kubo et.al.  have considered
the model where only one pair of Higgs doublets, coupled to the third
generation, acquires a non-vanishing v.e.v. In our case, it corresponds
to the following choice of the vectors $n_i$ and $\overline n_i$: $
n_i=\overline n_i = ( 0,0,1 )$, cf. eq.(18). For simplicity, they chose
the CKM and the Higgs mixing matrices equal to the unit matrix, i.e.
$K_{ij} = S_{ij} = \delta_{ij}$.  Using the two-loop renormalization
group equations for the gauge and Yukawa couplings with certain
boundary conditions and $m_{\tau}(M_Z), \alpha_{em}(M_Z)$ and
$\sin^2\theta_W(M_Z)$ as input, the authors of~\cite{Zoupanos} obtained
numerical values for $m_t(M_Z), m_b(M_Z)$ and $\alpha_s(M_Z)$ which are
in a good agreement with the experimental data. Masses of the
superpartners were assumed to be equal to $M_{SUSY}$.  The numerical
results were shown to depend rather weakly on the assumption that only
third generation of fermions becomes massive, that justifies the
approximation used in their analysis.

These calculations clearly demonstrate that the Finite model is
compatible with the experiment, though more complete analysis,
which takes into account the details of the soft supersymmetry breaking
is needed.

\vglue 0.5cm

\underline{Proton Decay}

The lower experimental limit on the proton lifetime is a very
rigid criterion for the viability of any GUT model. In the
supersymmetric unified theories, the proton decay takes place via
dimension-five operators that are generated due to the exchange of
heavy higgsino colour triplets.  In the minimal SUSY $SU(5)$ model, it
has been analyzed in Ref.\cite{AN}.  The preferred decay mode proves to
be $p \rightarrow \bar\nu K^+$ \cite{AN,AN2}. The amplitude of the
proton decay is proportional to:  \begin{equation} B_p \sim
 \frac{2\alpha_2}{\alpha_3 sin(2\beta)}\left(\frac{m_{\tilde{g}}}{m^2_{
\tilde{q}}}\right)\frac{3M_{GUT}}{M_{H_3}}10^6, \label{bp}
\end{equation}
where $M_{H_3}$ is the mass  of the heavy colour triplet. Experimentally one
has \cite{AN2}:
$$
B_p< (293\pm 42)\ Gev^{-1}.
$$
This constraint can be easily  satisfied for the low $\tan\beta$
 scenario, however, for the large $\tan\beta$ one, which is the case of
 the Finite model, one has problems due to the presence of the small
 factor of $\sin2\beta$ in the denominator of eq.(\ref{bp}).

In our model, besides the dimension-five operators analogous to that of
the minimal model which are generated by the  exchange of the fourth
colour triplet, there are additional ones that are generated by the
exchange of the three colour triplets adjusted to each generation. They
are mixed via the matrix $M_{ij}$.  To find their contribution, one has
to perform diagonalization by rotation of these three colour triplets
with the help of the same matrices $X$ and $\bar X$ that were used for
their doublet counterparts.  Then, proceeding along the lines of
Ref.\cite{AN2,AN3}, we will get the amplitude of the proton decay in
complete analogy with the minimal model:
\begin{equation} B_p \sim
 \frac{2\alpha_2}{\alpha_3 sin(2\beta)}\left(\frac{m_{\tilde{g}}}{m^2_{
\tilde{q}}}\right)3M_{GUT}\left(
\frac{1}{M^{(4)}_{H_3}}+2\frac{{\bar X}_{21}X_{21}}{M^{(1)}_{H_3}}+
2\frac{{\bar X}_{22}X_{22}}{M^{(2)}_{H_3}}+2\frac{{\bar X}_{23}X_{23}}{
M^{(3)}_{H_3}}\right)
10^6. \label{bp1}
\end{equation}
Since  the masses of all the colour triplets are of the same order of
magnitude, one can roughly put
$$ M^{(1)}_{H_3} \sim M^{(2)}_{H_3} \sim
M^{(3)}_{H_3} \sim M^{(4)}_{H_3} \sim 3 M_{GUT}, $$
and eq.(\ref{bp1}) becomes
$$
 B_p \sim \frac{2\alpha_2}{\alpha_3
sin(2\beta)}\left(\frac{m_{\tilde{g}}}{m^2_{
\tilde{q}}}\right)\left(1+2(\bar XX^T)_{22}\right)
10^6. \label{bp2}
$$
Now taking into account that the product of two orthogonal matrices can
always be written as $(\bar XX^T)_{22}=\cos\theta$, we get
\begin{equation}
 B_p \sim \frac{2\alpha_2}{\alpha_3
sin(2\beta)}\left(\frac{m_{\tilde{g}}}{m^2_{
\tilde{q}}}\right)\left(1+2\cos\theta\right)
10^6. \label{bp3}
\end{equation}
One can easily see from eq.(\ref{bp3}) that the additional factor
$(1+2\cos\theta)$ which we obtain in comparison to  the minimal
SUSY $SU(5)$ model can be used to
compensate for the smallness of $\sin 2\beta $ in the denominator and,
in this way we can avoid the problem with the proton decay in our model
in the case of large $\tan\beta $.

\vglue 0.5cm

\underline{FCNC}

The usual problem with the flavour changing neutral currents (FCNC)
in the  multihiggs models is that, due to the flavour mixing in the
Yukawa vertices of a general type, one cannot avoid the FCNC which is
already present at the tree level \cite{FCNC}. The other source of the
FCNC is the radiative corrections due to the Higgs mixing. Fortunately,
both the mechanisms of FCNC do not create any problem in our model. The
reason is that, first, the superpotential (\ref{YU}) is chosen in
a way that the Yukawa matrices are diagonal (or become diagonal after
the CKM rotation) in the generation space. This property of the
superpotential is not changed by the radiative corrections.  And,
second, since the finite model coincides with the MSSM below the GUT
scale, possible additional one-loop contributions to the FCNC,
different from the MSSM, are strongly suppressed. Thus, we face the
same problems with FCNC as in the MSSM.

After recent measurement of the branching fraction of the inclusive
decay $b \to s\gamma $~\cite{BR} special attention has been given
to this decay. The experimental value is very close to the prediction
of the SM which is given by the contribution of the so-called "penguin"
diagrams~\cite{pen}.  This means that an additional contribution from
SUSY  particles should be suppressed, which leads to a new constraint
on the parameters. The situation in the Finite model does not differ
>from the MSSM with large $\tan\beta$. One can meet the needed
requirement imposing  rather severe constraints on the soft breaking
terms. In this case the gluino contribution  may be essential.

\newpage

\section{Summary}

In conclusion, we summarize the main features of the model.
First, the requirement of a general ultraviolet finiteness of the
unified theory singles out almost a unique model and makes the theory
very rigid.  The Yukawa couplings appear to be polynomial functions in
terms of a unique $SU(5)$ gauge coupling.  Following Ref.~\cite{Pol},
we impose the universality conditions for the soft supersymmetry
breaking terms at the Planck scale and extend the requirement of
finiteness to them.  This requirement makes the number of free
parameters of the theory smaller than that of the minimal model. To
avoid the problem with the gauge couplings unification, which is usual
for the theory with enlarged Higgs sector \cite{dimop}, we reduce our
model to the MSSM below the GUT scale by the special fine-tuning
procedure. This fine-tuning is valid in the unified theory and does not
depend on the scale due to the finiteness of the latter, which makes
the choice of parameters more meaningful in our model. The low-energy
part of the theory, being the exact copy of the MSSM, bears an imprint
of the high-energy unified theory, thus resulting in the following
relations for the Yukawa couplings at the tree level:  $$ y^U_i=n_iy^U,
\ \ y^D_i=\bar n_iy^D, \ \ y^L_i=\bar n_iy^L.  $$ They reduce the
hierarchy of the Yukawa couplings  in the MSSM to the hierarchy of the
vacuum expectation values given by the projections  of the vectors
$n_i$ and $\bar n_i.$ These vectors, in their turn, are completely
defined by the Higgs sector of the unified theory, namely by the Higgs
mixing matrix.

Our main conclusion is that the finite supersymmetric Grand Unified
theory, being mathematically very rigid and unique, has passed all the
preliminary tests and is proved to be consistent. Being combined with
the soft supersymmetry breaking via supergravity, it naturally generates
the MSSM with some constraints on its parameters. The novel feature of
the model is the presence of additional heavy Higgs particles and
the mixing matrix in the Higgs sector which plays the crucial role in
the creation of the hierarchy of the Higgs field v.e.v's and via the
Higgs mechanism the hierarchy of quark and lepton masses.  Another
property of this matrix, which we do not discuss here, is its possible
contribution to the CP-violation due to the presence of phase factor in
$S$ analogous to that in the Kobayashi-Maskawa matrix.

\vspace{0.5cm}

{\bf Acknowledgements}

\vspace{0.5cm}
We are grateful to W.de Boer, G.Ross, S.Raby and G.Zoupanos for
useful discussions.

This work was partly supported by ISF grant No.RFL300 and by RFFI grant
No.94-02-03665-a. In addition the work of I.K. was supported by ICFPM
Fellowship (Grant INTAS No.93-2492).

\newpage

\section*{Appendix}
\setcounter{equation}{0}
\renewcommand{\theequation}{A\arabic{equation}}

Here we present the explicit  solution of the
minimization equations (\ref{mini}).  As one can see, eqs.(\ref{mini})
contain nonlinearity in the form of the quadratic combination $(\bar
v_k^2-v_k^2)$ which originates from the potential (\ref{10}). This is
the key property of the system which allows us to solve it
analytically.  As a first step, let us rewrite eqs.(\ref{mini}) in
the matrix form denoting this quadratic combination by $x$:
\begin{eqnarray} &&({\cal M}_1^2 +
xI)\overline{v} + {\cal M'} v = 0, \nonumber\\ &&({\cal M}_2^2 - xI)v +
{\cal M'}^{T} \overline{v} = 0, \label{A2}\\ &&x =
\frac{g^2+g^{\prime}{}^2}{4}\left(\overline{v}^2 - v^2 \right) ,
\nonumber
\end{eqnarray}
where $v~{\rm and}~\overline{v}$ are the real vectors in the
generation space
$$
\overline{v} =
\left(
\begin{array}{c}
\overline{v}_1 \\
\overline{v}_2 \\
\overline{v}_3 \\
\end{array}
\right),~~~
v =
\left(
\begin{array}{c}
v_1 \\
v_2 \\
v_3 \\
\end{array}
\right).$$
It is  obvious that if eqs.(\ref{A2}) have a nontrivial
solution, the condition
\begin{equation}
det\left(\begin{array}{cc}
{\cal M}_1^2 + xI & {\cal M'} \\
{\cal M'}^T & {\cal M}_2^2 - xI
\end{array}\right) = 0  \label{A3}
\end{equation}
should be satisfied.  Eq.(\ref{A3}) is the sixth order equation
with respect to $x$, but it can be easily factorized and solved.
Due to the diagonal structure of the matrices ${\cal M}_i$ and
${\cal M'}$, one has
$$
\left[(({\cal M}_1^2)_{11} +x)(({\cal M}_2^2)_{11} -x) - ({\cal
M'}_{11})^2 \right]\left[(({\cal M}_1^2)_{22} +x)(({\cal M}_2^2)_{22}
-x) -  ({\cal M'}_{22})^2 \right]
$$ $$
\left[(({\cal M}_1^2)_{33} +x)(({\cal M}_2^2)_{33} -x)
-({\cal M'}_{33})^2 \right] = 0, $$
which gives three solutions:
\begin{eqnarray}
x_i = \frac{1}{2}\left((m_2^2)_i - (m_1^2)_i \pm \sqrt{((m_1^2)_i +
(m_2^2)_i)^2 - 4(\mu_i)^2}\right),~~ i =1,~2,~3,\label{A31}
\end{eqnarray}
where we have introduced the notation similar to the MSSM:
$$ (m_1^2)_i =
({\cal M}_1^2)_{ii},~~~(m_2^2)_i = ({\cal M}_2^2)_{ii},~~~ \mu_i =
{\cal M'}_{ii}.$$

For each of the three $x_i$ given above, the system (\ref{A2}) is
factorized into three independent subsystems, but only one of them has
zero determinant and, consequently, nontrivial solution. Hence, there
exist three different independent solutions
\begin{eqnarray*} v_1 = \left(
\begin{array}{c}
u_1 \\
0 \\
0
\end{array}
\right), ~~
\overline v_1 &=&
\left(
\begin{array}{c}
{\overline u}_1 \\
0 \\
0
\end{array}\right);~~
v_2 = \left(
\begin{array}{c}
0 \\
u_2 \\
0
\end{array}
\right), ~~
\overline v_2 =
\left(
\begin{array}{c}
0 \\
{\overline u}_2 \\
0
\end{array}\right); \\
v_3 &=& \left(
\begin{array}{c}
0 \\
0 \\
u_3
\end{array}
\right), ~~
\overline v_3 =
\left(
\begin{array}{c}
0 \\
0 \\
{\overline u}_3
\end{array}\right).
\end{eqnarray*}
Here $u_i$ and  $\overline u_i$ are defined as
\begin{eqnarray}
u_i &=& \sqrt{\left((m_1^2)_i + (m_2^2)_i \pm \sqrt{((m_1^2)_i +
(m_2^2)_i)^2 - 4B^2\mu_i^2} \right)F_{\pm}(\mbox{\scriptsize
$B^2\mu_i^2$})}, \label{A4} \\
\bar u_i &=& -\mbox{
sign($B\mu_i$)}\sqrt{\left((m_1^2)_i + (m_2^2)_i \mp \sqrt{((m_1^2)_i +
(m_2^2)_i)^2 - 4B^2\mu_i^2}\right)F_{\pm}(\mbox{ $B^2\mu_i^2
$})}, \nonumber \\ \label{A5}
\end{eqnarray}
where
$$ F_{\pm }(B^2\mu_i^2)=\frac
1{g^2+g^{\prime 2}}\frac{\pm ((m_1^2)_i-(m_2^2)_i)-\sqrt{
((m_1^2)_i + (m_2^2)_i)^2-4B^2\mu_i^2}}{\sqrt{((m_1^2)_i + (m_2^2)_i)^2
-4B^2\mu_i^2}}. $$
The arbitrariness in the choice of the sign
in eqs. (\ref{A4}) and (\ref{A5}), originating from (\ref{A3}), is fixed
in the following way: we take the upper sign  if $(m_1^2)_i >
(m_2^2)_i$ and the lower sign in the opposite case.

The quantities $u_i$, $\bar u_i$ are real and positive by
definition. In order to get the right-hand sides of eqs. (\ref{A4}) and
(\ref{A5}) to be real and positive  and to have the potential bounded
>from below in the direction of vanishing quartic terms in (\ref{10}),
the following conditions should be satisfied:
\begin{eqnarray} &&
(m_1^2)_i + (m_2^2)_i > 2|B\mu_i|, \nonumber
\\ && (m_1^2)_i(m_2^2)_i < B^2\mu_i^2, \label{A7}
\end{eqnarray}
where the soft breaking parameter $B$ is of the order of $10^2-10^3$
GeV.

Due to our fine-tuning procedure, eqs.(\ref{m}),(\ref{f}), the
quantities $(m_1^2)_1$, $(m_1^2)_2$, $(m_2^2)_1$, $(m_2^2)_2$,
$\mu_1^2$ and $\mu_2^2$ are of the order of $M_{GUT}^2$, while
$(m_1^2)_3$, $(m_2^2)_3$ and $ \mu_3^2$ are of the order of $M_{Z}^2$.
Thus, eq.(\ref{A7}) can be satisfied for the third solution only. This
explains our choice of the vacuum solution.


\end{document}